\documentclass[11pt]{article}
\usepackage{graphicx}
\usepackage{rotating}
\usepackage{color}

\newcommand{\BABARPubYear}    {00}

\newcommand{\BABARConfNumber} {15}
\newcommand{\SLACPubNumber} {8537}

\usepackage{relsize}
\def\babar{\mbox{\slshape B\kern-0.1em{\smaller A}\kern-0.1em
    B\kern-0.1em{\smaller A\kern-0.2em R}}}

\def\epem       {\ensuremath{e^+e^-}}

\def\gaga  {\ensuremath{\gamma\gamma}}  

\def\qqbar {\ensuremath{q\overline q}}

\def\piz   {\ensuremath{\pi^0}}
\def\pizs  {\ensuremath{\pi^0\mbox\,\rm{s}}}

\def\pip   {\ensuremath{\pi^+}}
\def\pim   {\ensuremath{\pi^-}}

\def\pipm   {\ensuremath{\pi^\pm}}

\def\Kbar  {\kern 0.2em\overline{\kern -0.2em K}{}}

\def\Kpm   {\ensuremath{K^\pm}}
\def\Kp    {\ensuremath{K^+}}

\def\KS    {\ensuremath{K^0_{\scriptscriptstyle S}}} 
\def\KL    {\ensuremath{K^0_{\scriptscriptstyle L}}} 
\def\Kstarz  {\ensuremath{K^{*0}}}

\def\Kstar   {\ensuremath{K^*}}

\def\Kzb   {\ensuremath{\Kbar^0}}
\def\KzKzb {\ensuremath{K^0 \kern -0.16em \Kzb}}

\def\Dbar  {\kern 0.2em\overline{\kern -0.2em D}{}}

\def\Dzb   {\ensuremath{\Dbar^0}}
\def\DzDzb {\ensuremath{D^0 {\kern -0.16em \Dzb}}}

\def\Bz    {\ensuremath{B^0}}
\def\B     {\ensuremath{B}}
\def\Bbar  {\kern 0.18em\overline{\kern -0.18em B}{}}

\def\Bzb   {\ensuremath{\Bbar^0}}
\def\Bu    {\ensuremath{B^+}}

\def\BB    {\ensuremath{B\Bbar}} 
\def\BzBzb {\ensuremath{B^0 {\kern -0.16em \Bzb}}}

\def\jpsi  {\ensuremath{{J\mskip -3mu/\mskip -2mu\psi\mskip 2mu}}} 
\mathchardef\Upsilon="7107
\def\Y#1S{\ensuremath{\Upsilon{(#1S)}}}

\def\FourS {\Y4S}

\mathchardef\Deltares="7101
\mathchardef\Xi="7104
\mathchardef\Lambda="7103
\mathchardef\Sigma="7106
\mathchardef\Omega="710A
\def\Deltabar   {\kern 0.25em\overline{\kern -0.25em \Deltares}{}}
\def\Lbar {\kern 0.2em\overline{\kern -0.2em\Lambda\kern 0.05em}\kern-0.05em{}}
\def\Sigbar{\kern 0.2em\overline{\kern -0.2em \Sigma}{}}
\def\Xibar{\kern 0.2em\overline{\kern -0.2em \Xi}{}}
\def\Obar{\kern 0.2em\overline{\kern -0.2em \Omega}{}}
\def\Nbar{\kern 0.2em\overline{\kern -0.2em N}{}}
\def\Xbar{\kern 0.2em\overline{\kern -0.2em X}{}}

\def\mes        {\mbox{$m_{\rm ES}$}}
\def\mec        {\mbox{$m_{\rm EC}$}}

\def\ev   {\ensuremath{\rm \,e\kern -0.08em V}}
\def\kev  {\ensuremath{\rm \,ke\kern -0.08em V}} 
\def\mev  {\ensuremath{\rm \,Me\kern -0.08em V}} 
\def\gev  {\ensuremath{\rm \,Ge\kern -0.08em V}} 
\def\gevc {\ensuremath{{\rm \,Ge\kern -0.08em V\!/}c}} 
\def\tev  {\ensuremath{\rm \,Te\kern -0.08em V}}
\def\mevc {\ensuremath{{\rm \,Me\kern -0.08em V\!/}c}} 
\def\gevcc{\ensuremath{{\rm \,Ge\kern -0.08em V\!/}c^2}} 
\def\mevcc{\ensuremath{{\rm \,Me\kern -0.08em V\!/}c^2}}

\def\mm   {\ensuremath{\rm \,mm}}

\def\invfb   {\ensuremath{\mbox{\,fb}^{-1}}}
\def\mus  {\ensuremath{\rm \,\mus}}

\def\mus        {\ensuremath{\,\mu{\rm s}}}    
\def\mrad{\ensuremath{\rm \,mr}}                

\def\calA{{\ensuremath{\cal A}}}

\def\gsim{{~\raise.15em\hbox{$>$}\kern-.85em
          \lower.35em\hbox{$\sim$}~}}
\def\lsim{{~\raise.15em\hbox{$<$}\kern-.85em
          \lower.35em\hbox{$\sim$}~}}

\def\CP                 {\ensuremath{C\!P}}
\def\ra                 {\ensuremath{\rightarrow}}
\def\to                 {\ensuremath{\rightarrow}}

\def\pep2{PEP-II}

\newcommand{\etapr}{\ensuremath{\eta^{\prime}}}

\newcommand{\etaprrg}{\ensuremath{\etapr\ra\rho^0\gamma}}

\newcommand{\dedx}{\ensuremath{\mathrm{d}\hspace{-0.1em}E/\mathrm{d}x}}

\newcommand{\eqref}[1]{Eq.~(\ref{eq:#1})}

\newcommand{\epjc}      [1]  {{Eur.\ Phys.\ Jour.\ C~{\bf #1}}}

\newcommand{\pl}        [1]  {{Phys.\ Lett.\ {\bf #1}}}      

\newcommand{\prl}       [1]  {{Phys.\ Rev.\ Lett.\ {\bf #1}}} 

\def\jetset74   {\mbox{\tt Jetset \hspace{-0.5em}7.\hspace{-0.2em}4}}

\newcommand{\calB}{\mbox{${\cal B}$}}

\def\dbline{\noalign{\vskip 0.15truecm\hrule}\noalign{\vskip 2pt}\noalign{\hrule\vskip 0.15truecm}}
\def\sgline{\noalign{\vskip 0.20truecm\hrule\vskip 0.20truecm}}

\newcommand{\dzpi}{$\Bu \to \Dzb \pip, \Dzb \to \Kp \pim$}

\newcommand{\kstpi}{$\Bu \to \Kstarz \pip$}
\newcommand{\rhok}{$\Bu \to \rho^0 \Kp$}
\newcommand{\rhozpipm}{\mbox{$\Bu \to \rho^0 \pip$}}
\newcommand{\ppp}{$\Bu \to \pip \pim \pip$}
\newcommand{\kpp}{$\Bu \to \Kp \pim \pip$}

\newcommand{\rhoppim}{\mbox{$\Bz \to \rho^{\mp} \pi^{\pm}$}}

\newcommand{\etapKp}{\mbox{$\Bu \to \etapr \Kp$}}

\newcommand{\etapKz}{\mbox{$\Bz \to \etapr K^0$}}
\newcommand{\etapKs}{\mbox{$\Bz \to \etapr \KS$}}

\newcommand{\omegah}{\mbox{$\Bu \to \omega h^+$}}

\newcommand{\omegapi}{\mbox{$\Bu \to \omega \pip$}}

\newcommand{\omegaKp}{\mbox{$\Bu \to \omega \Kp$}}

\newcommand{\omegaKz}{\mbox{$\Bz \to \omega K^0$}}
\newcommand{\omegaKs}{\mbox{$\Bz \to \omega \KS$}}

\newcommand{\etaprd}{\ensuremath{\etapr\ra\eta\pi^+\pi^-}}

\newcommand{\DE}{\ensuremath{\Delta E}}
\newcommand{\de}{\ensuremath{\Delta E}}

\newcommand{\xf}{\mbox{${\cal F}$}}

\newcommand{\kzs}{\mbox{$\KS$}}

\def\ome{\mbox{${\omega}$}}
\def\trpi{\pim\pip\piz}

\newcommand{\kst}{\mbox{$K^{*0}$}}

\newcommand{\pvec}{{\bf p}}
\newcommand{\half}{\mbox{${1\over2}$}}

\usepackage{epsf}  
\usepackage{epsfig}  
\newcommand{\psfile}[3][]{ 
  \begin{center}
    \setlength{\epsfxsize}{#3\linewidth}\leavevmode
    \def\noOpt{}\def\testit{#1}\ifx\testit\noOpt%
      \epsfbox{#2}%
    \else%
      \epsfbox[#1]{#2}%
    \fi
  \end{center} 
}

\newcommand{\psfiletwoBB}[5]{ 
  \begin{minipage}{\linewidth}
    \parbox[b]{.49\linewidth}{%
      \begin{center}
        \setlength{\epsfxsize}{#5\linewidth}\leavevmode\epsfbox[#1]{#2}
      \end{center}
    }
    \hfill
    \parbox[b]{.49\linewidth}{%
      \begin{center}
        \setlength{\epsfxsize}{#5\linewidth}\leavevmode\epsfbox[#3]{#4}
      \end{center}
    }
  \end{minipage}
}
\newcommand{\sigGSBratio}{\calA}

\long\def\inst#1{\par\nobreak\kern 4pt\nobreak
    {\it #1}\par\vskip 10pt plus 3pt minus 3pt}

\begin{document}
{\pagestyle{empty}

\begin{flushright}
\babar-CONF-\BABARPubYear/\BABARConfNumber \\
SLAC-PUB-\SLACPubNumber
\end{flushright}

\par\vskip 3cm

\begin{center}
\Large \bf Measurements of charmless three-body \\ 
       \bf and quasi-two-body {\boldmath $B$} decays
\end{center}
\bigskip

\begin{center}
\large The \babar\ Collaboration\\
\mbox{ }\\
July 25, 2000
\end{center}
\bigskip \bigskip

\begin{center}
\large \bf Abstract
\end{center}
We present preliminary results of a search for several exclusive
charmless hadronic $B$ decays from electron-positron annihilation data 
collected by the \babar\ detector near the \FourS\ resonance. 
These include three-body decay modes with final states 
$h^{\pm}h^{\mp}h^{\pm}$ and $h^{\pm}h^{\mp}\pi^0$, and quasi-two-body 
decay modes with final states $X^0 h$ and $X^0 \KS$, where $h = \pi$ or 
$K$ and $X^0 = \etapr$ or $\omega$.
We find $\calB(\rhoppim) = (49\pm 13^{+6}_{-5})\times 10^{-6}$ and
$\calB(\etapKp) = (62 \pm 18 \pm 8)\times 10^{-6}$
and present upper limits for eight other decays.

\vfill
\begin{center}
Submitted to the XXX$^{th}$ International 
Conference on High Energy Physics, Osaka, Japan.
\end{center}

\newpage
}

\begin{center}
\small

The \babar\ Collaboration
\bigskip

B.~Aubert,
A.~Boucham,
D.~Boutigny,
I.~De Bonis,
J.~Favier,
J.-M.~Gaillard,
F.~Galeazzi,
A.~Jeremie,
Y.~Karyotakis,
J.~P.~Lees,
P.~Robbe,
V.~Tisserand,
K.~Zachariadou
\inst{Lab de Phys.\ des Particules, F-74941 Annecy-le-Vieux, CEDEX, France}
% ...

\end{center}\newpage

\setcounter{footnote}{0}

\section{Introduction}
\label{sec:Introduction}
The charmless hadronic decays of neutral $B$ mesons of interest in this 
paper can be used to explore \CP\ violation arising from two possible 
interference effects.  Indirect \CP\ violation in \Bz\ decays can arise from
interference between the direct amplitude and one involving \Bz--\Bzb\
mixing, while direct \CP\ violation in charged $B$ decays can occur via 
interference between tree and penguin amplitudes.  The former
offers the future possibility 
of measuring directly the CKM angle $\alpha$ of the Standard Model. It is 
expected~\cite{ref:physbook} that this will require samples of order 
100 million $B$ mesons.  The magnitude of direct \CP\ violation is more
difficult to estimate but interesting constraints can be provided with
much smaller samples.
In addition, the tree-penguin interference may allow measurement of
the CKM angle $\gamma$ \cite{ref:gammeas}, via measurements of the decay
rates of modes presented here and the related $B\ra K\pi$ and
$B\ra\pi\pi$ decay modes \cite{ref:conf14}.

The initial challenge is to obtain significant samples of these rare decays.
Many of the modes of interest have only 
recently been observed for the first time, or remain undiscovered 
\cite{ref:CLEOjun00,ref:CLEOetap,ref:PDG2k}. In this paper, we describe
preliminary searches of  
the initial \babar\ data sample for a number of charmless hadronic $B$ decays,
and give preliminary measurements of their branching fractions.
The three-body final states are limited to those with
at least two charged tracks (\pipm\ or \Kpm) and at most one \piz.
The hadronic resonances $\rho$ and \Kstar\ are sufficiently 
short-lived that the modes which contain them interfere quantum mechanically 
with other relevant 3-body final state modes. We have however searched
for them independently using appropriate kinematic selections within the 
Dalitz plot. The quasi-two-body decays 
involve an $\etapr$ or $\omega$ meson accompanied by a neutral kaon or
charged pion or kaon. We summarize the decay modes 
considered\footnote{Charge conjugate decay modes are 
assumed throughout this paper.} as follows:

\vskip 0.5cm

\begin{center}
\begin{tabular}{lcl}

 three-body                                    &	& quasi-two-body \cr
\sgline
	\kstpi					& 	& \etapKp \cr
	\rhok					& 	& \etapKs \cr
	\kpp					& 	& \omegapi \cr
	\rhozpipm				& 	& \omegaKp \cr
	\ppp					& 	& \omegaKs \cr
	\rhoppim				&	& \cr

\end{tabular}
\end{center}

\vskip 0.3cm

\section{The \babar\ detector and data}
\label{sec:babar}
The data used in the analyses were collected with the \babar\
detector at the \pep2\ storage ring. The \babar\ detector, described in detail 
elsewhere~\cite{ref:detectorPaper}, consists of five active
sub-detectors. Surrounding the beam-pipe is a silicon vertex
tracker (SVT) to track particles of momentum less than $\sim$120\mevc\ and
to provide precision measurements of the positions of charged
particles of all momenta as they leave the interaction point. A beam-support
tube surrounds the SVT. Outside this is a 40-layer drift
chamber (DCH), filled with an 80:20 helium-isobutane gas mixture to
minimize multiple scattering, providing measurements of track momenta
in a 1.5 T magnetic field.  It also provides \dedx\ measurements to help 
charged particle identification. Surrounding the outer
circumference of the drift chamber is a novel detector of internally
reflected Cherenkov radiation (DIRC) which provides charged hadron
identification in the barrel region. This consists of quartz bars of
refractive index $\sim$1.42 in which Cherenkov light is produced by
relativistic charged particles. This is internally reflected and
collected by an array of photomultiplier tubes, which enable Cherenkov
rings to be reconstructed and associated with the charged tracks in
the DCH, providing a measurement of particle velocities. Outside
the DIRC is a CsI(Tl) electromagnetic calorimeter (EMC) which is used to
detect photons and neutral hadrons, and to provide electron identification. 
The EMC is surrounded by a
superconducting coil which provides the magnetic field for
tracking. Outside the coil, the flux return is instrumented with
resistive plate chambers (IFR), interspersed with iron which may be
used for the identification of muons and \KL\ mesons.

The data sample used for the analyses contains 8.8~million 
$\BB$ pairs~\cite{ref:detectorPaper}, 
corresponding to 7.7\invfb\ taken
on the \Y4S\ resonance. In addition, 1.2\invfb\ of data taken
off-resonance have been
used to validate the contribution to backgrounds resulting from $e^+e^-$
annihilation into light \qqbar\ pairs. These data have all been processed 
with reconstruction software to determine
the three-momenta and positions of charged tracks and the energies and
positions of photons and merged \pizs. 
Refined information on particle type from the various sub-detectors described 
above is also provided, and is used in particle identification algorithms in 
the analyses, as described below.

\section{Candidate selection}
\label{sec:Selection}
Charged tracks are required to satisfy some standard track criteria,
including a requirement that the momentum is less than 10\gevc\ and that 
the transverse momentum is greater than 0.1\gevc. They are required to have 
at least 20 hits in the DCH and to originate close to the beam-spot.

Photon candidates are identified in our calorimeter as deposits of energy 
unassociated with charged tracks. Resolved \pizs\ are reconstructed by 
combining pairs of photon candidates and requiring that the invariant mass
of the resultant candidate is between 100\mevcc\ and 170\mevcc. 
Photon candidates used in \piz\ reconstruction are required to have a 
minimum energy of 50\mev. In addition to resolved \pizs, unresolved 
\pizs\ are identified as single clusters of energy deposited in the EMC 
without distinct local maxima and satisfying certain topological criteria 
\cite{ref:detectorPaper}. For the final event
selection, we tighten the requirement on the \piz\ mass to between 
120\mevcc\ and 150\mevcc.

Neutral kaons are reconstructed through the decay $\KS \to \pip \pim$.
\kzs\ candidates are formed from combinations of two
oppositely charged tracks satisfying basic track criteria similar to 
those mentioned above, except with a looser constraint on their proximity 
to the beam spot. We require that the \kzs\ candidate's flight length 
exceed 2\mm, the angle between its flight and momentum directions 
be less than 40\mrad, and the mass lie within $\pm10$\mevcc\ of nominal.

\begin{figure}[ht]
\begin{center}
\mbox{\epsfig{figure=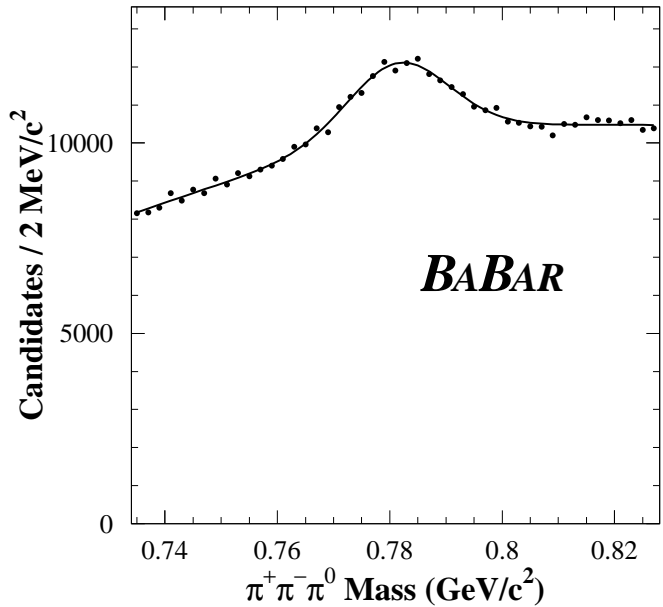,height=3.5in}}
\end{center}
\caption{Mass spectrum of hard $\omega$ candidates in on-resonance data, 
fitted with a Breit-Wigner convoluted with a Gaussian
 for the signal and a second-order polynomial for the background.}
\label{fig:resom}
\end{figure}

We reconstruct $\ome$ mesons using the dominant decay channel, $\ome\ra\trpi$,
which has a branching fraction of 88.8\%.
Candidates are obtained as combinations of two charged tracks with 
opposite signs and one $\piz$ candidate. The invariant mass of the 
$\ome$ candidate is required to be within 0.05\gevcc\ of the known 
$\omega$ mass~\cite{ref:PDG2k}.
Figure~\ref{fig:resom} shows the $\trpi$ invariant mass for $\ome$
candidates with a center-of-mass momentum between 1.9 and 3.1\gevc.
The signal is fitted with a Breit-Wigner function with the natural width of
the $\ome$ (8.4\mevcc) convoluted with a Gaussian resolution function,
and the combinatorial background is fitted with a second-order polynomial.
The Gaussian resolution function is found to have a width of 8\mevcc.

\begin{figure}[ht]
\psfiletwoBB{65 85 535 755}{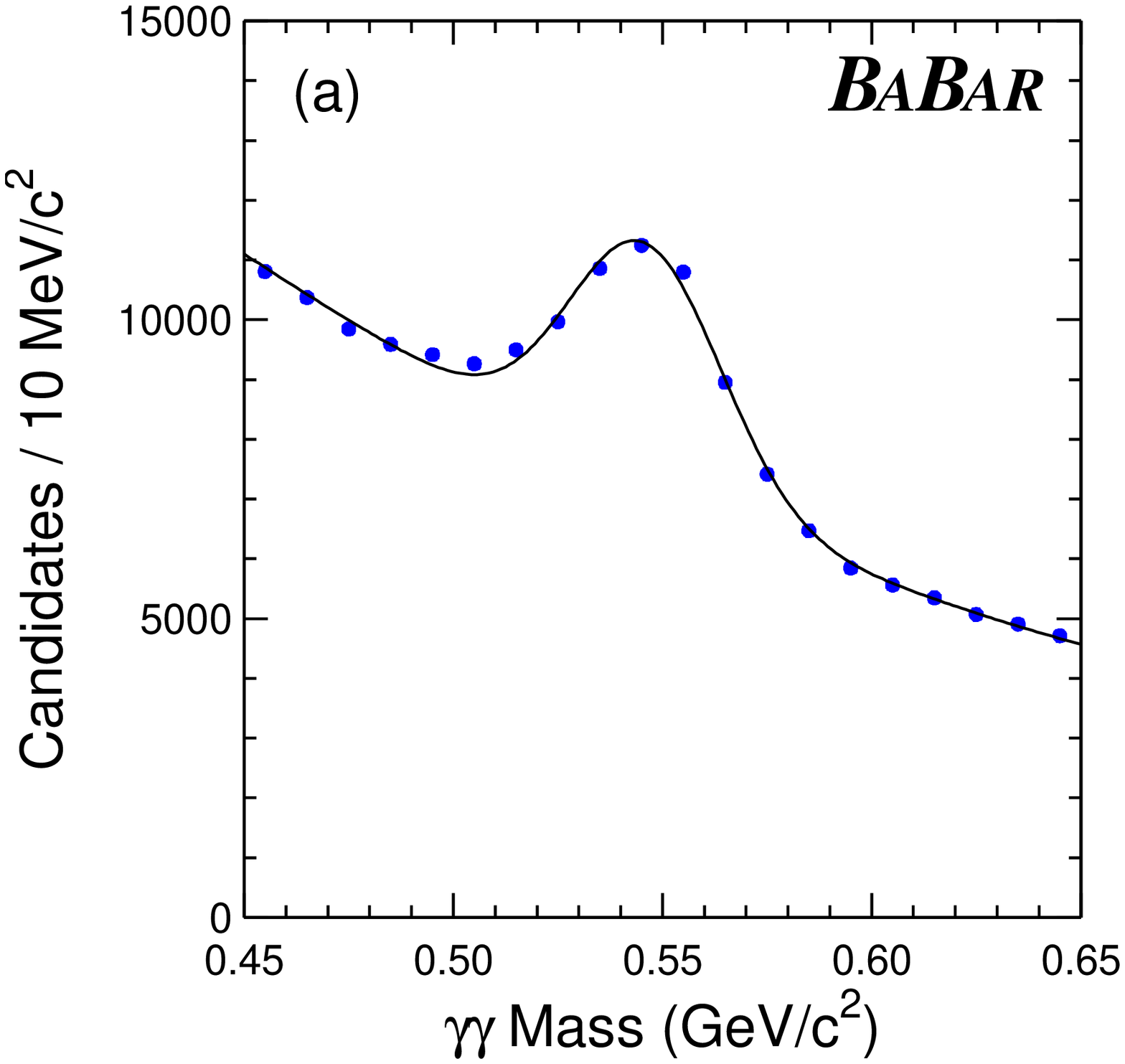}{65 85 535 755}{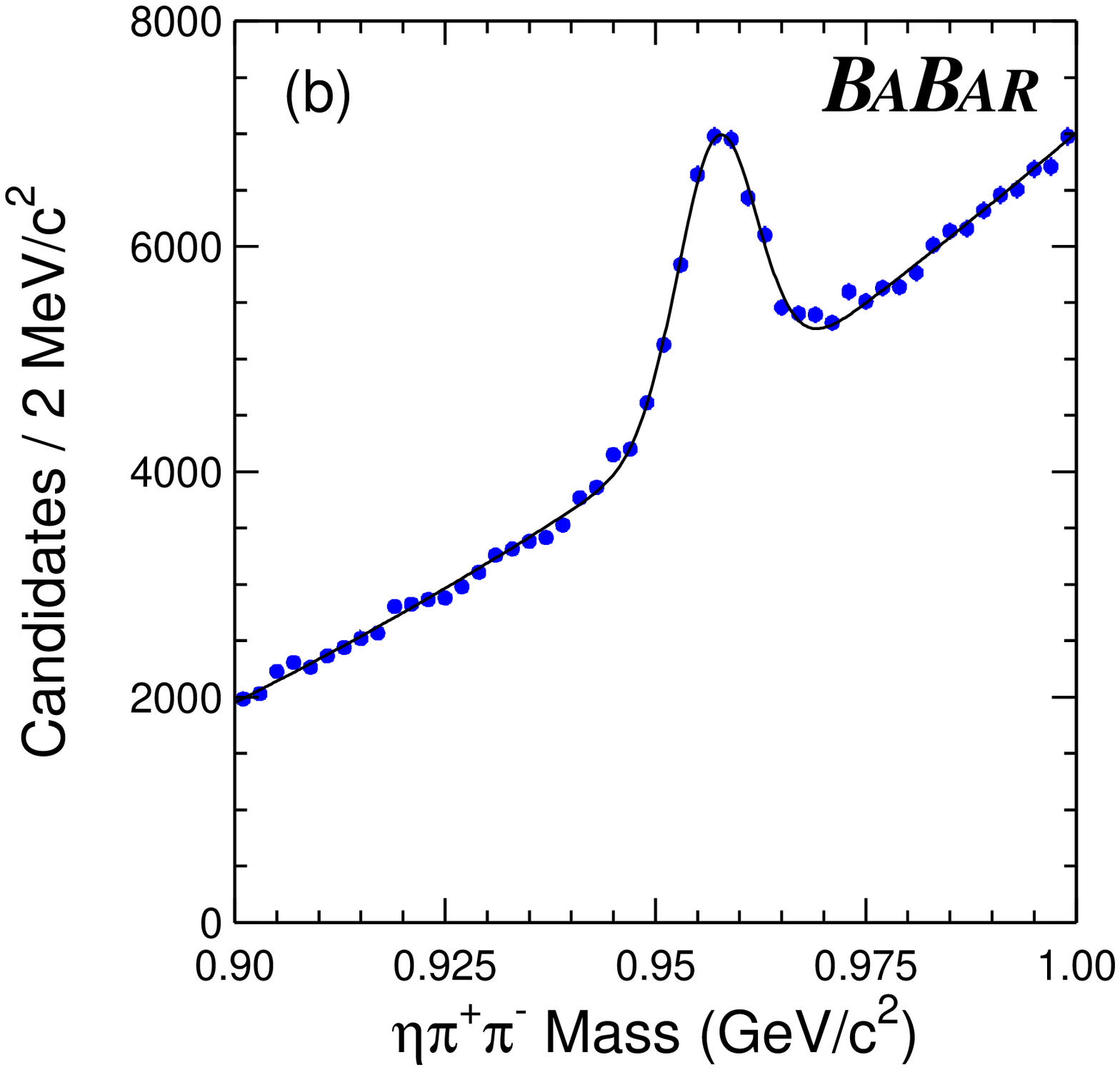}{.98}
 \vspace{-0.3cm}
 \caption{\label{fig:etapEppMass}%
Distribution of (a) $\gamma \gamma\ (\sigma=19\ \mevcc)$ mass, and
(b) $\eta\pi^+\pi^-\ (\sigma=4.7\ \mevcc)$ mass in on-resonance data.} 
\end{figure}

Candidate $\rho$ and $\Kstar$ mesons are reconstructed by combining pairs of 
appropriately charged tracks and/or \piz\ candidates, with the assumption 
of the relevant final-state rest masses, and requiring that the combination 
have an invariant mass sufficiently close to the known resonance mass. 
Requirements for the invariant masses are analysis-dependent,
and are described below.

The $\eta$ meson is reconstructed by combining pairs of photon candidates 
with a minimum energy of 50\mev. The reconstructed $\eta$ mass for
data is shown in Fig.~\ref{fig:etapEppMass}. We obtain a width of 19\mevcc\ 
for a Gaussian fit, with the background described by a 
second-order polynomial.
The \etapr\ meson is reconstructed in the \etaprd\ decay mode. 
This is done by combining $\eta$ candidates with pairs of oppositely 
charged tracks. A mass constraint 
is used to improve the $\eta$ mass resolution.
Fig.~\ref{fig:etapEppMass} shows the invariant
mass distribution for all $\eta\pi\pi$ candidates with momentum above 
2\gevc.  From a fit to the sum of a Gaussian and a second-order
polynomial, we obtain a width of 4.7\mevcc\ for the \etapr\ peak.

For $B$ decays to final states formed from a pseudoscalar and a vector
meson, the vector meson is polarized. We make use of the
angular distribution of the vector meson decay products to distinguish 
between signal and background. We compute in the rest frame of the vector
meson the cosine of the angle $\theta_H$ between the
direction of one of the vector meson daughters (the normal to the decay
plane in the case of the $\omega$) and the direction of the $B$ meson in the same 
frame. We expect signal events to be distributed
according to $\cos^2{\theta_H}$, while background should be approximately flat.

To veto electrons, we require that charged hadrons satisfy $E/p <
0.9$, where $E$ is the energy measured in the calorimeter associated
with a charged track of momentum $p$. We also require that kaons be
positively identified as such with a combination of the information
from the DIRC and the drift chamber \dedx. For pions, we simply
require that they not be identified as kaons.

Reconstruction of $B$ candidates is done by forming all combinations of
the appropriate final-state candidate particles, and requiring them
to satisfy kinematic constraints appropriate for $B$ mesons. 
We use two kinematic variables~\cite{ref:detectorPaper} for this: 
$\mes = \sqrt{(\half s + \pvec_0\cdot \pvec_B)^2/E_0^2 - p_B^2}$,
where the
subscripts $0$ and $B$ refer to the \epem\ system and the $B$ candidate,
respectively; and
$\Delta E = E_B^* - \sqrt{s}/2$, where $E_B^*$ is the \B\ candidate
energy in the center-of-mass frame.
For signal events, the former has a value close to the $B$ meson mass
\footnote{ A nearly equivalent constrained mass that we use for some of the 
modes is $\mec = \sqrt{{E_B^*}^2 - {p_B^*}^2}$,
obtained from a fit constrained by $E_B^* = \half\sqrt{s}$.}
and the latter should be close to zero.
In our analyses, the appropriate final state masses are
assigned to the final state particles in calculating \DE.

We have measured the expected distributions of the signal with respect to 
\mes\ and \DE, using two calibration modes described in 
Section \ref{sec:Validation}.
Based on these, we define a ``signal region'' in the \mes--\DE\ plane by 
$|\mes-\left<\mes\right>|<6$\mevcc\ and $|\DE-\left<\DE\right>|<70$\mev, where 
$\left<\mes\right>$ and $\left<\DE\right>$ indicate the mean values obtained in the 
calibration modes.
This signal region is common to all modes, except \omegah\ (see Section 
\ref{q2results}).

\section{\boldmath Calibration with charmed \B\ decays}
\label{sec:Validation}
The decay mode \dzpi\
results in a final state containing three charged tracks, while the same
$\Bu$ decay with $\Dzb \to \Kp \pim \piz$ yields in addition a \piz.
The $D^0$~mass is small compared to that of the~$B$, so that 
this decay has kinematics similar to charmless
$B$~decays to two light mesons such as $B \rightarrow \rho\pi$,
$B \rightarrow K^* \pi$ or $B\ra\omega\pi$.  
The product branching fractions are known to about 10\% precision and are
roughly an order of magnitude larger than those expected of many 
charmless modes. A study of these decays can therefore be used to
calibrate the variables used in 
the charmless analyses. The most significant difference 
is that the~$D^0$ has a measurable lifetime, resulting in decay tracks which 
do not share a common vertex with the pion from the~$B$ decay.

The candidate selection proceeds along similar lines to those 
described earlier for the charmless decay modes, except that
the invariant mass of the $D^0$ daughters, assumed to be $\pi K
(\piz)$, must be consistent with the~$D^0$ mass. The background is 
estimated by counting events in a ``sideband'' of the \mes--\DE\ distribution, 
and also by looking
at $K\pi\pi(\piz)$ track combinations which satisfy the selection criteria 
but which have an incorrect charge combination (the kaon charge is
opposite to that of the $B$~meson). The extrapolation from sideband to
signal region is done in the same way as described for the charmless
analyses.

The efficiency of the selection for \dzpi[\piz]
is $(23 \pm 2)\%$ [$(12\pm1)\%$], evaluated as described in Section
\ref{sec:Physics}.
When combined with the total number of $B$~decays the
product branching fraction for 
\dzpi[\piz] is measured to be
$(2.21 \pm 0.11) \times 10^{-4}$ [$(6.79 \pm 0.32) \times
10^{-4}$],  
where the errors are statistical only. 
These are both consistent with the corresponding world-average
values~\cite{ref:PDG2k}.

The selected signals in these modes have been used to
estimate the \mes\ and 
\DE\ resolution functions of our target modes in data.
We find the \mes\ resolution to be $2.7 \pm 0.1\ \mevcc$ in the mode
with only three charged tracks
and $3.0 \pm 0.1$~\mevcc\ in the mode with a \piz.
The corresponding \DE\ resolutions are $26\pm 2$~\mev\ in the mode
with only three charged tracks
and $33 \pm 2$~\mev\ in the mode with a \piz.
The measured values were used to inform the choice of signal region, 
noted at the end of Section \ref{sec:Selection} and for calculation of the 
selection efficiency and its uncertainty.

\section{Background characterization and suppression}
\label{sec:background}
Charmless hadronic modes suffer very large amounts
of background from random combinations of tracks, mostly from light quark
and charm continuum production. Such backgrounds may be reduced by 
selection requirements on the event topologies computed in the 
\FourS\ rest frame. We use the angle $\theta_T$ between the
thrust axis of the $B$ meson decay and the thrust axis of the rest of the 
event. For continuum-related backgrounds, these two directions
tend to be aligned because the reconstructed $B$ candidate daughters
generally lie in the same jets as those in the rest of the event.
By contrast, in $B$ events, the decay products from one $B$ meson are 
independent of those in the other, making the distribution of this angle
isotropic. In consequence,
requiring that this opening angle be significant provides a strong 
suppression of continuum backgrounds.


Other event shape variables also help to separate signal and background. 
We combine several variables in a Fisher discriminant \xf\ \cite{Fisher}.
The variables contained in \xf\ are:
\begin{itemize}

\item The summed energy in nine annular cones of angular extent 175 mrad, 
coaxial with the thrust axis of the $B$ candidate.

\item The absolute value of the cosine of the angle between the $B$ direction
and the beam axis.

\item The absolute value of the cosine of the angle between the thrust axis of
the $B$ candidate and the beam axis.

\end{itemize}
The Fisher discriminant is a linear combination of these 11 variables. The
coefficients for each variable are chosen to maximize the
separation between training samples of signal and background events.
These coefficients were determined for the \omegah\ analysis and the
same coefficients are used for the other $\omega$ and $\etapr$ analyses.

Despite the power of such topological variables to reduce the combinatorial 
backgrounds, most of the modes we have searched for continue to suffer 
significant background levels. Even after stringent selection criteria have 
been applied, it is necessary to do a background subtraction to isolate 
a signal or set an upper limit. 
In order to do this, the background in the signal region must be estimated. 
This is done by noting that the amount of background in the signal region 
should be related to the amount in a sideband region, located 
near the signal region in the \mes-\DE\ plane. We define 
\sigGSBratio, to be the ratio of the number of candidates in the 
signal region to the number in the sideband region. Two different methods 
have been used to estimate \sigGSBratio. In the first, the shape of the 
distribution of the background as a function of \mes\ is measured from 
on-resonance data. This is done with events slightly displaced from the 
signal region in the \DE\ variable ($0.1 < |\DE\ | < 0.3$). 
In the second method, off-resonance data are used, 
simply counting the numbers of candidates in the signal and sideband
regions to provide the ratio.
The first method is used in the three-body measurements, and the second
method in the quasi-two-body measurements. In each case, the alternative
method has been used to evaluate systematic errors. 
Where insufficient data are available
for the off-resonance studies, selection criteria in uncorrelated
variables have been loosened, to provide additional statistics in the 
determination of \sigGSBratio. 
We find that the value of \sigGSBratio\ is quite independent of the
decay mode provided the selection criteria are common. For some modes with
limited statistics we substitute a more precise value of \sigGSBratio\ from a
mode in which it is better measured.
Simulated light-quark ($u$, $d$, $s$, $c$) continuum events
are used to check the value of \sigGSBratio\
and provide a statistically-independent sample for 
optimization of event selections in the final stage of the analysis.

\section{Analysis}
\label{sec:Physics}
The branching fractions are calculated according to 
\begin{equation}
{\cal B} = \frac{N_1-\sigGSBratio N_2}{N_{\BB} \times \epsilon}
\label{BReqn}
\end{equation}
where $N_1$ is the number of candidates in the signal region for
on-resonance data; $N_2$ is the number of candidates in on-resonance data 
observed in the sideband region, so that $\sigGSBratio N_2$ is the 
estimated number of background candidates in the signal region; 
$N_{\BB}$ is the number of 
$\BB$ pairs produced and $\epsilon$ is the signal efficiency.
Except in particular cases, explained below, the sideband is specified by 
$5.20 < \mes < 5.27$\gevcc, $|\DE-\left<\DE\right>|<0.2$\gev, 
where $\left<\DE\right>$ was defined in Section \ref{sec:Selection}.

The numbers and distributions of candidates within the signal region 
remain unknown to us until all aspects of the analysis are finalized. The final 
selection criteria are chosen to maximize the sensitivity of the signal, 
defined as the expected signal yield divided by its statistical uncertainty. 
Once chosen, neither the background description nor the cuts are changed.
This procedure has been followed independently for each channel.

For the signal efficiency in Eq.~(\ref{BReqn}), we use simulated signal
events and the same selection criteria as used for the data. The
efficiencies due to tracking, particle identification and the \de\ and 
\mes\ selection criteria are determined from independent control samples
from the data.

\subsection{Modes with three-body final states}
\label{3bresults}
The selection criteria which are varied during optimization are the 
thrust-angle requirement, the helicity-angle requirement and the mass
requirement on resonance candidates.  In addition to these and the selection
criteria described in Section \ref{sec:Selection}, a quality requirement of 
$\chi^2 < 20$ is applied to the three (two) charged track vertices in all 
charged (neutral) three-body analyses. Finally, a veto on all pairs
of tracks which are consistent with the $\Dzb \ra K^+\pi^-$ decay hypothesis
(independent of particle identification) is applied in all three-body analyses.
This has no significant effect on our results in any mode except \rhok.
The results for these analyses are summarized in Table~\ref{tab:3bodytab}.
The branching fraction central values are calculated from Eq.\ \ref{BReqn}.  
The significance values given are the probabilities of a fluctuation of
the background to account for the observed yield, in equivalent Gaussian
standard deviations.

\begin{table}[!htbp]
\color[rgb]{1,1,1}
\caption{\relax}
\normalcolor
\label{tab:3bodytab}

\begin{center}

\begin{sideways}
\makebox[8.25in]{

\begin{tabular}{lcccccc}
\multicolumn{7}{c}{Table 1: Results for the three-body final state analyses. Energies and masses are in units of~\mev. See the text for an explanation} \cr
\multicolumn{7}{l}{of the quantities given.} \cr\cr

\dbline

Quantity			& \kstpi & \rhok  & \kpp   &\rhozpipm& \ppp   &\rhoppim \cr

\sgline

$\cos \theta_T$			&  0.70  &  0.60  &  0.70  &  0.55  &  0.70   &  0.50   \cr
$\kst$/$\rho$ mass		&$\pm 100$&$\pm 200$&N/A&$\pm 200$&N/A&$\pm 150$ \cr
$\kst$/$\rho$ helicity angle	&  0.40  &  0.40  &  N/A   &  0.40  &  N/A    &  0.30   \cr
Signal events			&	 &	  &	   &	    &	      &		\cr
~~On-res data			&   22   &   25   &   33   &   64   &   32    &   77	\cr
~~Off-res data			&    1   &    1   &    3   &    4   &    5    &    6	\cr
Sideband events			&	 &	  &	   &	    &	      &		\cr
~~On-res data			&  319   &  256   &  454   & 1057   &  722    & 1129	\cr
~~Off-res data			&   50   &   45   &   60   &  166   &   92    &  183	\cr
${\cal A}$			& 0.037  & 0.056  & 0.037  & 0.037  & 0.037   & 0.037	\cr
Est. BG				&$11.8\pm0.9$&$14.3\pm1.2$&$16.7\pm1.1$&$39.1\pm1.9$&$26.6\pm1.4$&$41.5\pm4.3$ \cr
Signal				&$10.2\pm4.8$&$10.7\pm5.1$&$16.3\pm5.8$&$24.9\pm8.2$&$5.4\pm5.7$&$35.5\pm9.8$ \cr
$\epsilon\times~{\rm secondary}~\calB_i$  (\%)
			&$9.7\pm0.9$&$9.8\pm1.0$&$6.0\pm0.6$&$11.9\pm1.2$&$8.2\pm0.8$&$8.3^{+1.0}_{-1.1}$ \cr

\sgline

Stat. sign. ($\sigma$)		&   2.4  &   2.2  &   3.2  &   3.3  &    0.7  &    4.5	\cr
cross-talk			&	 &	  &	   &	    &	      &		\cr
~~correction ($\times10^{-6}$)	&	 & $-2.0$ &	   &	    &	      &		\cr
\calB ($\times10^{-6}$) &$13\pm6\pm1$&$10\pm6\pm2$&$31\pm11\pm3$&$24\pm8\pm3$&$7.5^{+7.9}_{-7.5}\pm0.8$&$49\pm13^{+6}_{-5}$ \cr
~~90\%\ CL limit		& $<28$& $<29$& $<54$& $<39$ & $< 22$ &		\cr

\dbline

\end{tabular}

}
\end{sideways}

\end{center}

\end{table}

The background extrapolation is validated, albeit with low statistics, 
by the observed ratios of events in signal and sideband regions in
off-resonance data. In each case, we found that the number of events falling 
in the signal region was consistent within one standard deviation with the 
expectation from the sideband regions.

The \rhok\ mode has the additional complication of cross-talk background 
from the \etapKp\ mode with \etaprrg.  The $\gamma$ can be quite soft 
and in such cases the $\rho^0$ and $\Kp$ may imitate the \rhok\ signal.
We correct for this by processing simulated events from this 
background mode using the full set of optimized selection criteria for 
the \rhok\ mode. 
We obtain an efficiency for this mode of 2.4\%, leading to a reduction 
of the observed branching fraction by $2.0\times 10^{-6}$. The existence
of this $B$-related background was foreseen, and for this reason the sideband
used to estimate the background is somewhat smaller than that used 
for the other modes, to ensure that there was no significant leakage of 
this background into the light-quark background estimate.
Cross-talk among the other modes considered is found from Monte Carlo 
simulation to be negligible at the current level of precision.

\begin{figure}[!htbp]
 \psfile{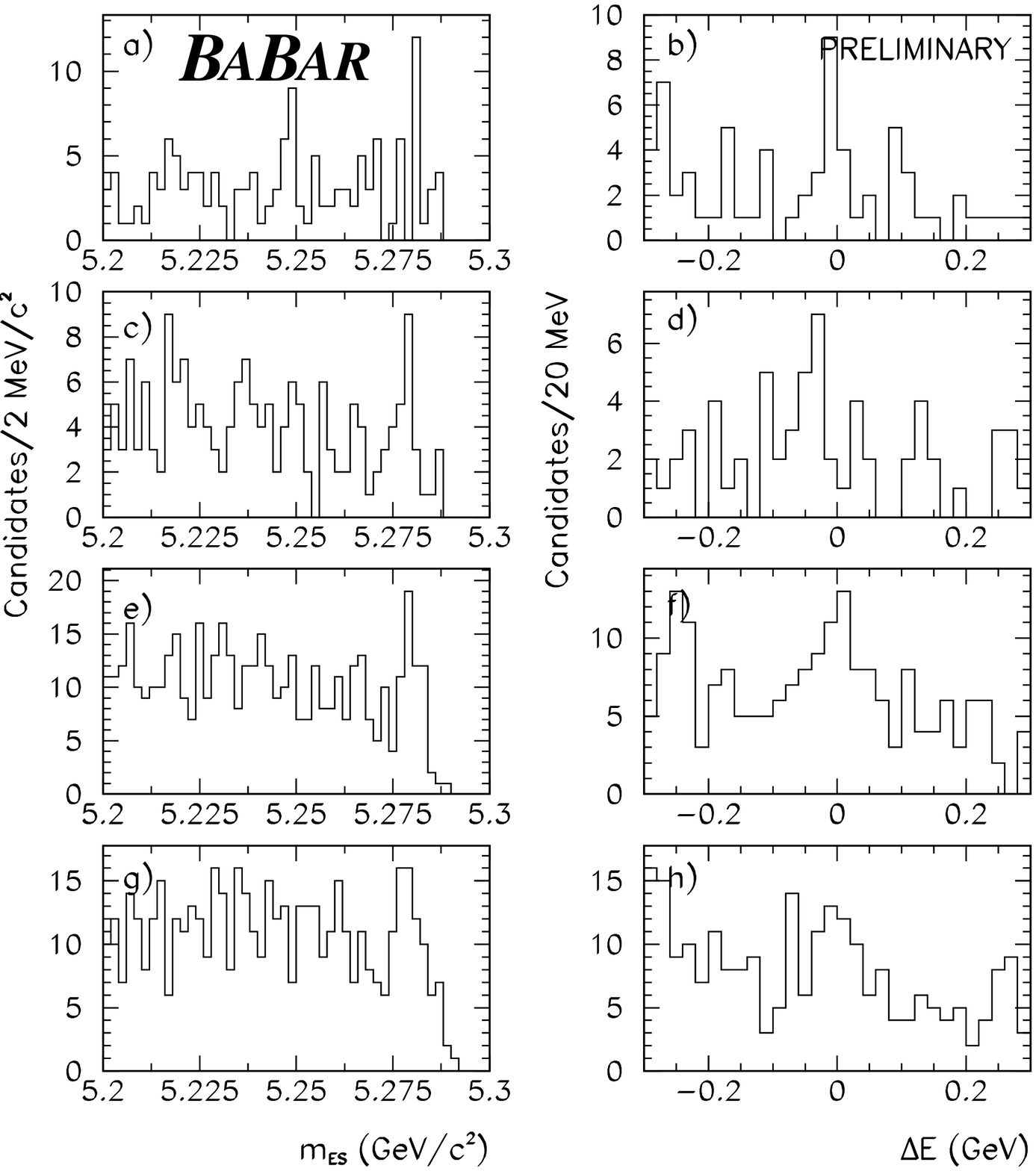}{0.9}
 \vspace{-0.3cm}
 \caption{\label{fig:3bodyplots}%
 Distributions in \mes\ and \DE\ after all other selection criteria have been 
applied for (a,~b)~\kstpi\ , (c,~d)~\rhok\ , (e,~f)~\rhozpipm\ , 
(g,~h)~\rhoppim\ .}
\label{resmodes} 
\end{figure}

The distributions of \mes\ and \DE\ for the resonant three-body modes
are shown in Fig.~\ref{resmodes}.

\begin{figure}[!htb]
\begin{center}
\mbox{\epsfig{file=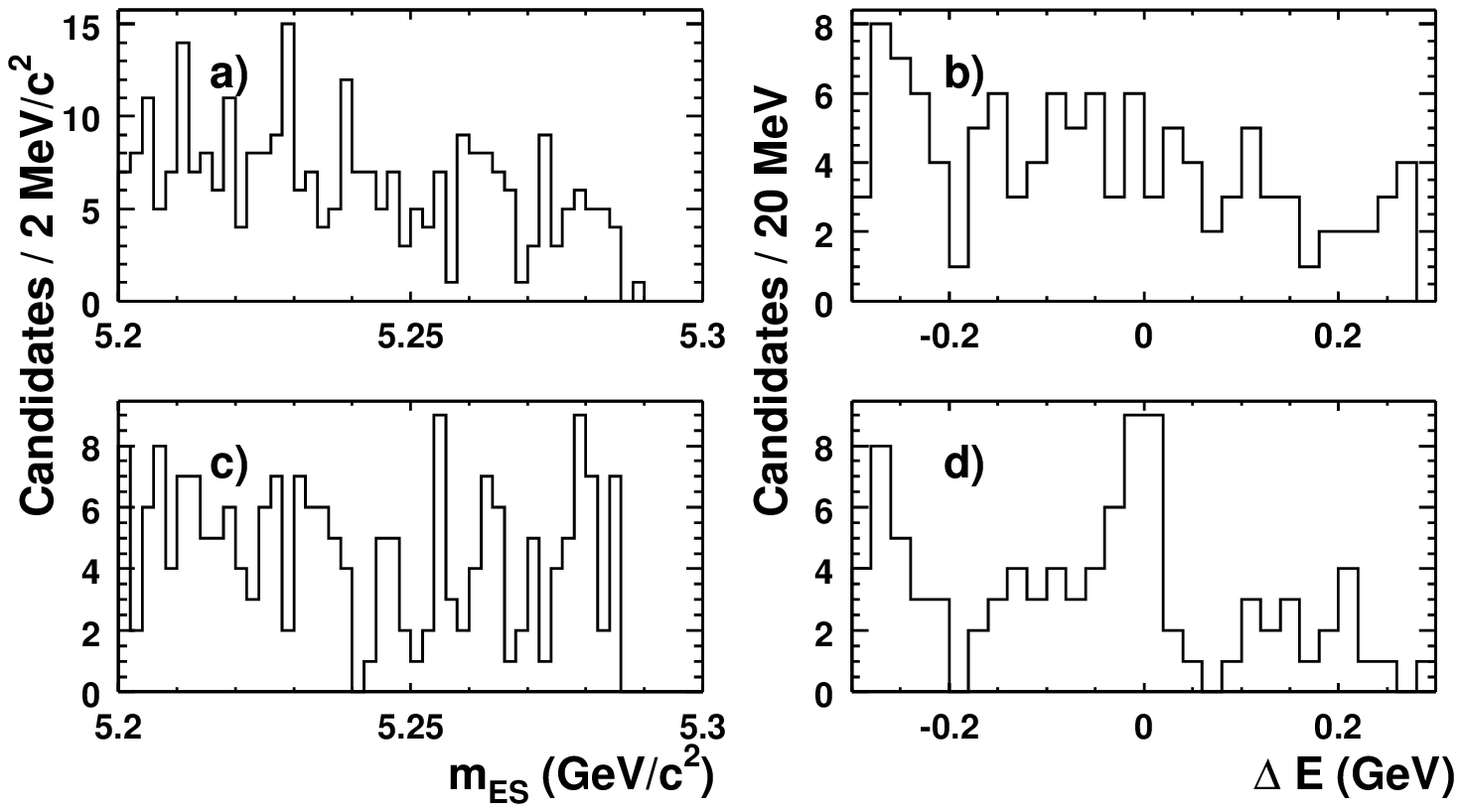,width=12cm}}
\end{center}
\caption[\mes\ and \DE\ distributions for \ppp\ and \kpp ]
{\mes\ and \DE\ distributions for (a,~b)~\ppp\ and (c,~d)~\kpp\ 
non-resonant analyses after the final selection.}
\label{NRmodes}
\end{figure}

We note that the analyses of the two modes \kpp\ and \ppp\ differ
in some respects from those with a resonance. We remove all two-body
combinations with invariant masses less than 2~$\gevcc$,
and in addition veto combinations of tracks consistent with the 
decay mode $\Bu \to \jpsi \Kp$. The distributions of \mes\ and
\DE\ for the non-resonant modes are shown in Fig.~\ref{NRmodes}.

\subsection{\boldmath Modes with $\omega$ and \etapr}
\label{q2results}
We summarize the results of the measurements in Table~\ref{tab:numtab}.  For
all decay modes, we require $|\cos \theta_T|<0.9$, and then find the optimum
selection requirement in \xf. The latter is listed for each mode in the 
table. Also listed are the selection criteria on resonance masses and 
$\pi^0$ or $\eta$ masses. The selection criteria are centered on the known 
masses~\cite{ref:PDG2k}. We also require that $|\cos{\theta_H}|$ be greater 
than 0.55 (0.4) for the decay modes \omegah(\omegaKz).

Table~\ref{tab:numtab} gives the yields in the signal and sideband regions,
both for on- and off-resonance data. The signal region is defined as $|\mes
-5.279| < 0.006$ GeV/$c^2$ and $|\DE| < 0.070$ GeV, except for \omegah\ where
we use $-0.113 < \DE < 0.070$ GeV, allowing for the 43 MeV shift of \omegaKp\
when the energy of the $K^+$ is computed with a pion mass. The sideband is
defined as $5.20 < \mes < 5.27$ GeV/$c^2$ and $|\DE| < 0.2$ GeV. 
The branching fraction central values and significance values are
calculated in the same way as for Table~\ref{tab:3bodytab}. 

\begin{table}[!htbp]
\caption{
Results for the quasi-two-body decay analyses. Energies and masses are in units of~\mev.
See the text for an explanation of the quantities given.
}
\label{tab:numtab}
\begin{small}
\begin{center}
\vskip 0.2cm
\begin{tabular}{lcccc}
\dbline
Quantity & \omegah &\omegaKz &\etapKp & \etapKz \cr
\sgline
\xf\   & $-0.4$ & $-0.2$  & 0.3   &   0.6 \cr
$\omega$/$\etapr$ mass& $\pm$20 &$\pm$20&$\pm$10 &$\pm$10\cr
$\piz$/$\eta$ mass    & $\pm$15 &$\pm$15&$\pm$45 &$\pm$10\cr
Signal events         &      &      &       &      \cr
~~On-res data         &  13  &  0   &  14   &   2  \cr
~~Off-res data        &   1  &  0   &   0   &   0  \cr
Sideband events       &      &      &       &      \cr
~~On-res data         & 128  &  19  &  45   &  14  \cr
~~Off-res data        &  16  &  6   &   8   &   1  \cr
${\cal A}$& 0.055 &0.042& 0.042 & 0.042\cr
Est. BG               & $7.1\pm0.9$ & $0.8\pm0.2$ & $1.9\pm0.3$ & $0.6\pm0.2$
 \cr
Signal		      & $5.9\pm3.6$ & $-0.8\pm0.0$ & $12.1\pm3.7$ & $1.4\pm1.4$ \cr
Eff. $\epsilon$ (\%)  & 8.5 &  7.6 & 17.1  & 10.0 \cr
Secondary $\calB$ (\%)& 88.8 &30.5&17.2& 6.0 \cr
$\epsilon\times~{\rm secondary}~\calB$  (\%)    
                      & $7.5\pm1.4$ & $2.3\pm0.4$ & $2.9\pm0.4$
                      & $0.60\pm0.09$ \cr
\sgline
Stat. sign. ($\sigma$)&  1.7 & 0.0   & 5.3 &  1.1 \cr
\calB ($\times10^{-6}$)
                      & $8.9 \pm 5.4 \pm 2.2$ & 0.0 & $62 \pm 18 \pm 8$
                      & $27 \pm 27 \pm 5$ \cr
~~90\%\ CL limit      & $<24$& $<14$ &       &$<112$\cr
\dbline
\end{tabular}
\end{center}
\end{small}
\end{table}

\begin{figure}[!htbp]
 \psfile[60 385 530 605]{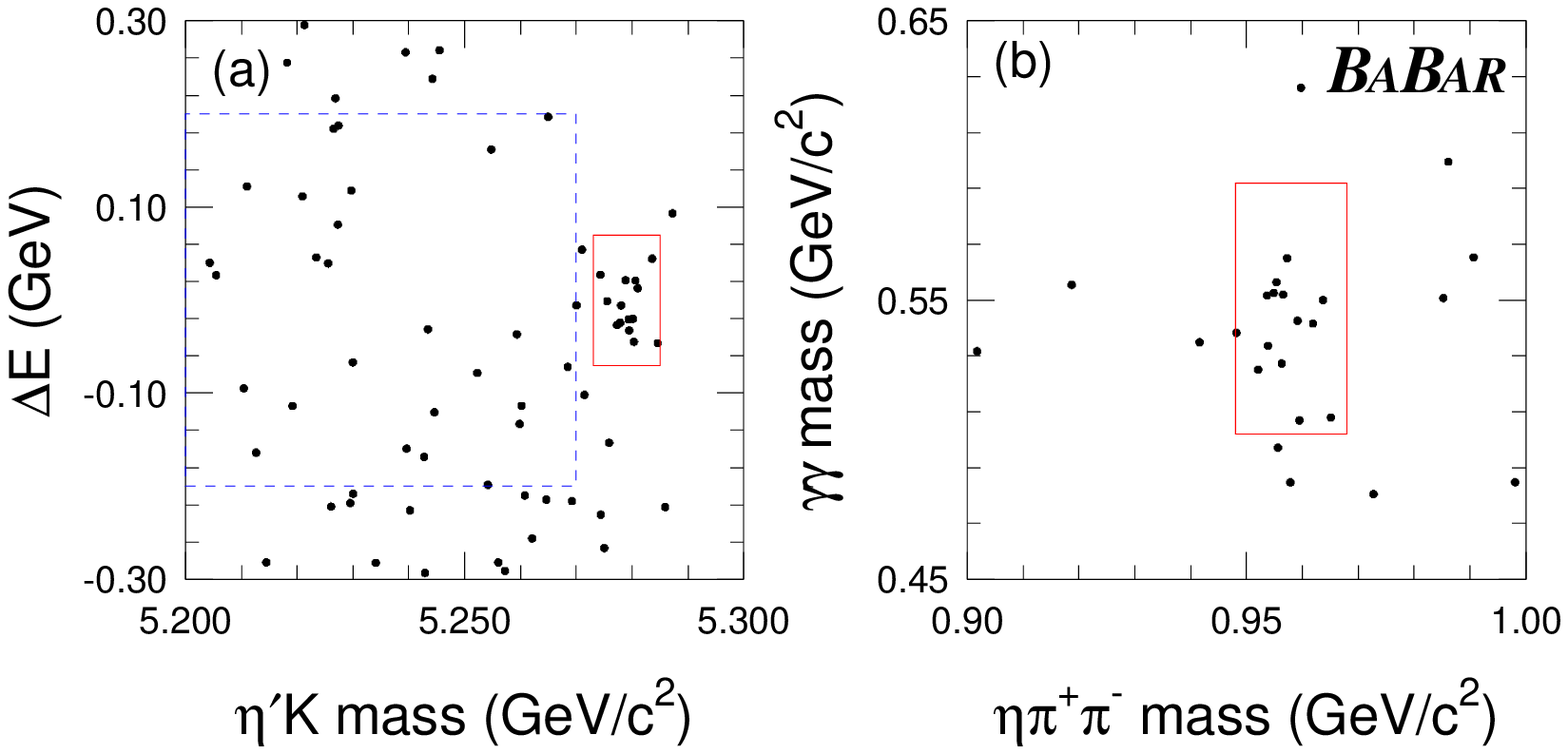}{0.8}
 \vspace{-0.3cm}
 \caption{\label{fig:etaprk_2d.ps}%
 Kinematics of \etapKp, \etaprd:  (a)~\DE\ vs.~\mes\ and 
(b)~\gaga\ vs.~$\eta\pi^+\pi^-$ invariant mass.
 }  
\end{figure}

In Fig.\ \ref{fig:etaprk_2d.ps}
we illustrate the reconstruction of the \etapKp\ states with plots
of the
kinematic variables associated with the $B$ meson and daughter resonance.  For
the other decay modes we report our final results as 90\% confidence upper
limits.

\begin{figure}[!htbp]
 \psfile[65 160 560 650]{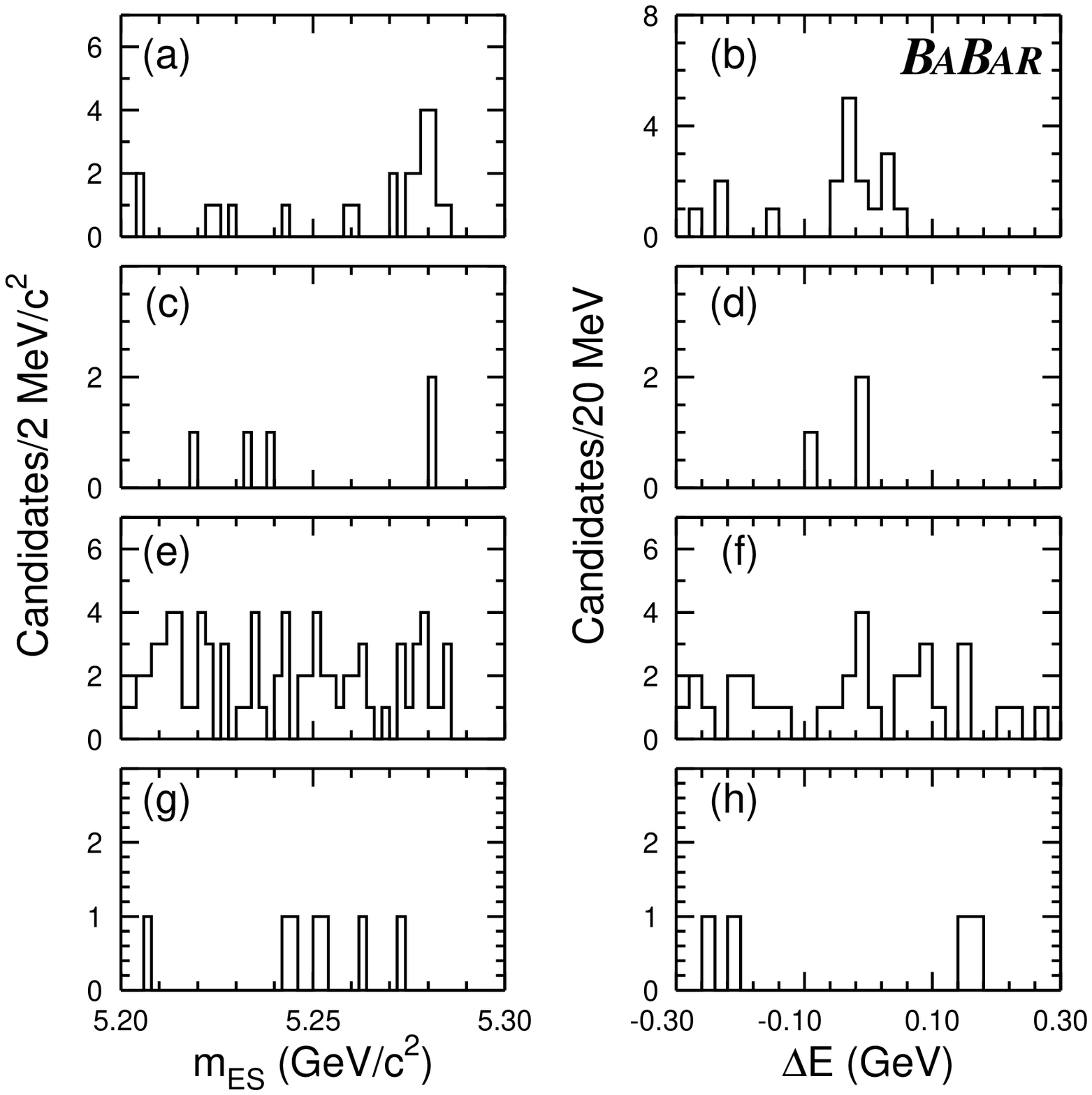}{0.8}
 \vspace{-0.3cm}
 \caption{\label{fig:q2bmbdeProj}%
 Distributions in \mes\ and \DE\ after all other selection criteria have been 
applied for (a,~b)~\etapKp\ , (c,~d)~\etapKz\ , 
(e,~f)~\omegah\ , (g,~h)~\omegaKz\ .}  
\end{figure}

The distributions for all decay modes, projected on to the \mes\ and \DE\ axes,
are presented in Fig.\ \ref{fig:q2bmbdeProj}.

\section{Systematic errors}
\label{sec:Systematics}

The primary uncertainty in the signal yield is the counting statistics
of the accepted events in the signal region. The systematic uncertainty on the background
estimate receives contributions from the number of events observed in the
on-resonance sideband region and from the ratio \sigGSBratio. The uncertainty
on the latter depends on the statistics of the off-resonance samples and on
the method used to determine it. We estimate an overall uncertainty in 
\sigGSBratio\ of 4\% for the charged three-body modes and 10\% for
the other modes.

The branching fraction itself has a multiplicative uncertainty from our
knowledge of the efficiency, $\epsilon$. 
We measure
$\epsilon$ using signal Monte Carlo simulation, and a significant uncertainty
arises from limited signal mode statistics, the fractional error due to
the size of the Monte Carlo samples used to estimate $\epsilon$
varies between 4\% and 7\%, depending on the mode.  The accuracy of
the  simulation is subject to systematic uncertainties in the efficiency
of tracking and calorimetric shower detection, particle
identification criteria, and the resolutions on quantities used to define
selection criteria.  At present the dominant
uncertainties are in the tracking, $\pi^0/\eta$ reconstruction, and Fisher
Discriminant selection efficiency, with some variation on their relative 
importance depending on the decay mode.

Independent studies determine an absolute track-finding
efficiency uncertainty of 2.5\% per track, added
coherently for all tracks required in the analyses. For $\pi^0$ or $\eta$
reconstruction, we assign an uncertainty of 5\%, and add an additional
uncertainty of 1-5\% in quadrature to account for the effect of the final
mass requirement. For the Fisher Discriminant selection, the uncertainty 
on the efficiency can vary considerably, depending on the tightness of the
selection requirement. The uncertainty ranges from only a few percent for 
decay modes with low background, to as much as 15\% for ones which need 
a tight selection requirement. 
The number of produced \BB\ events has been estimated in separate
studies \cite{ref:detectorPaper} with an estimated uncertainty of 3.6\%.
Uncertainties in the 
particle identification and \mes\ and \DE\ selection criteria
contribute systematic uncertainties of 2-4\%, depending on mode.
The overall
systematic uncertainty is the quadrature sum of the contributions from all
sources.

For modes in which we find an upper limit for the branching fraction, we
account for the uncertainty of the efficiency by using in Eq.\
\ref{BReqn}\ the measured efficiency reduced by one standard
deviation in the systematic error.

\section{Summary}
\label{sec:Summary}
We have presented a number of preliminary measurements 
of charmless hadronic $B$ decays, summarized in Table \ref{tab:results}.
We observe significant signals for \etapKp\ and \rhoppim, 
with branching fractions
$\calB(\etapKp) = (62 \pm 18 \pm 8)\times10^{-6}$ and
$\calB(\rhoppim) = (49\pm13^{+6}_{-5})\times10^{-6}$.
Upper limits are given for the remaining channels we have studied.
In each case, our determinations are consistent with earlier 
published measurements \cite{ref:CLEOjun00,ref:CLEOetap,ref:PDG2k}.

\vskip 0.5cm

\begin{table}[!htbp]
\caption{
Summary of branching fraction measurements.  Inequality denotes 90\% CL
upper limit, including systematic uncertainties.
}
\label{tab:results}
\begin{center}
\begin{tabular}{lc}
\dbline
Decay mode		& \calB ($\times10^{-6}$) \cr
\sgline
	\kstpi		& $<28$          \cr
	\rhok		& $<29$	         \cr
	\kpp		& $<54$	         \cr
	\rhozpipm	& $<39$         \cr
	\ppp		& $<22$	         \cr
	\rhoppim	& $49\pm13^{+6}_{-5}$ \cr
        \etapKp         & $62 \pm 18 \pm 8$   \cr     
	\etapKz         & $<112$         \cr 
	\omegah         & $<24$          \cr
	\omegaKz        & $<14$          \cr
\dbline
\end{tabular}
\end{center}
\end{table}

\vskip 1.0cm

\section*{Acknowledgments}
\label{sec:Acknowledgments}


We are grateful for the contributions of our \pep2\ colleagues in
achieving the excellent luminosity and machine conditions
that have made this work possible.
We acknowledge support from the
Natural Sciences and Engineering Research Council (Canada),
Institute of High Energy Physics (China),
Commissariat \`a l'Energie Atomique and
Institut National de Physique Nucl\'eaire et de Physique des Particules
(France),
Bundesministerium f\"ur Bildung und Forschung
(Germany),
Istituto Nazionale di Fisica Nucleare (Italy),
The Research Council of Norway,
Ministry of Science and Technology of the Russian Federation,
Particle Physics and Astronomy Research Council (United Kingdom), the
Department of Energy (US),
and the National Science Foundation (US). In addition, individual support 
has been received from the Swiss 
National Foundation, the A. P. Sloan Foundation, the Research Corporation,
and the Alexander von Humboldt Foundation.
The visiting groups wish to thank 
SLAC for the support and kind hospitality
extended to them.

\end{document}